\documentclass[twocolumn,preprintnumbers,amsmath,amssymb,prl]{revtex4-1}

\usepackage{amsmath,amssymb,epsfig,graphicx}

\usepackage{dcolumn}

\def\be{\begin{eqnarray}}
\def\ee{\end{eqnarray}}
\def\bea{\begin{eqnarray}}
\def\eea{\end{eqnarray}}

\begin{document}

\preprint{}

\title{Between symmetry and duality in supersymmetric QFT
}

\author{Shlomo S. Razamat,$^1$ Orr Sela,$^1$ and Gabi Zafrir$^2$
}

\affiliation{
$^1$Department of Physics, Technion, Haifa 32000, Israel\\ $
^2$Kavli IPMU (WPI), UTIAS, the University of Tokyo, Kashiwa, Chiba 277-82583, Japan\\
}

\date{\today}

\begin{abstract}
We  study two cases of interrelations  between enhancement of symmetries in the infra red (IR) and duality properties of supersymmetric quantum field theories in four dimensions. 
First we  discuss an $SU(2)$ ${\cal N}=1$ model with four flavors, singlet fields, and a superpotential. We show that this model flows to a conformal field theory with $E_6\times U(1)$ global symmetry. 
The enhancement of the flavor symmetry follows from Seiberg duality.
The second example is concerned with an $SU(4)$ gauge theory with matter in the fundamental and antisymmetric representations. We  argue that this model has enhanced $SO(12)$ symmetry in the IR, and then guided by this enhancement, we deduce a new IR duality.

\end{abstract}


\maketitle

%\end{CJK*}

\section{Introduction}
Models with different ultra violet (UV) properties can flow to the same IR conformal fixed point.  In supersymmetric setups there are many examples of such universality properties with the  UV models being gauge theories having different gauge groups and gauge singlets, or having the same gauge groups  with  different gauge singlet fields and different superpotentials. In the latter case the phenomenon is usually referred to as self duality.  The global symmetry of the two dual models will usually act differently on the gauge non singlet fields.

Another interesting  phenomenon is that of the global symmetry in the IR being larger than the symmetry in the UV.  This often happens when some of the degrees of freedom become  free as one approaches the fixed point. However, there are also cases when the symmetry enhancement happens as the quantum numbers of the states at the IR fixed point align to form representations of a bigger symmetry. The bigger symmetry usually will  have the same rank as the symmetry in the UV but larger dimension. Enhancements of the rank are also possible though will not be discussed here. 

We will discuss in this short note two cases where self duality of a certain model can be related to enhancement of symmetry in a similar model.  The basic observation is that in the case of self duality one often can add additional gauge singlet operators on the two sides of the duality, without spoiling the IR equivalence, such that the two dual models will have exactly the same field content. The duality will still identify the symmetries of the two models in a non obvious way leading to symmetry enhancement. It will be also observed in one of the examples that in order to obtain a model with enhanced symmetry the additional gauge singlet fields will break some of the original symmetry.

We will emphasize three important guiding principles. First, breaking global symmetries with interactions might lead to an enhanced symmetry in the IR which is not a subgroup of,  in fact might not contain, the symmetry of the original model. Second, self-dualities of field theories can be utilized to find theories with enhanced flavor symmetry by constructing models which are  structurally invariant under dualities with the effect of the latter being a non trivial action on the matter. Third,  enhancement of symmetry can in some cases be a sign of new self dualities.

\section{$E_6$  symmetry from duality} 
\noindent{\bf The $E_6$ model:}
Let us consider $SU(2)_g$ gauge theory with eight fundamental chiral fields. We split the eight chiral fields  into six ($Q_A$) and two ($Q_B$). We also introduce gauge invariant operators $M_A$ and $M_B$ coupling as,
\be
M_A Q_A Q_A+M_B Q_B Q_B\,  .
\ee The quiver theory is depicted in Fig. 1 and charges of fields can be obtained in the table below. The choice of gauge singlet fields breaks the symmetry of the model from $SU(8)$ down to $SU(6)_A\times SU(2)_B\times U(1)_h$. We will soon show that this enhances to $E_6\times U(1)_h$. We also note that $SU(8)$ is not subgroup of $E_6$, and for the enhancement to be possible it is crucial to break the symmetry.

\begin{figure}[h]
\includegraphics[scale=0.46]{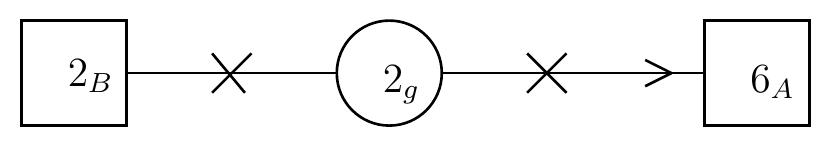}\caption{Model with $E_6\times U(1)$ global symmetry. The cross on the edges denotes the fields $M_A$ and $M_B$. These are flipping the baryonic operators constructed from the fields associated to the corresponding edge. Flipping chiral operator ${\cal O}$ means introducing chiral field $\phi_{\cal O}$ and coupling it as $\phi_{\cal O}{\cal O}$.}
\end{figure}

\

\begin{tabular}{|c||c|c|c|c|c|}
Field&$SU(2)_g$&$SU(2)_B$&$SU(6)_A$&$U(1)_h$&$U(1)_{\hat r}$\\ \hline
 $Q_A$&${\bf 2}$&${\bf 1}$ &${\bf 6}$&$\,\;\frac1{2}$& $\frac5{9}$ \\
$Q_B$&${\bf 2}$&${\bf 2}$&${\bf 1}$&$-\frac3{2}$&$\frac13$\\ 
$M_B$&${\bf 1}$&${\bf 1}$&${\bf 1}$&$\,\;3$&$\frac43$\\ 
$M_A$&${\bf 1}$&${\bf 1}$&$\overline {\bf 15}$&$-1$&$\frac{8}{9}$\\

\hline
\end{tabular}

\

In the table $U(1)_{\hat r}$ is the superconformal R symmetry obtained by a maximization \cite{Intriligator:2003j}  and the conformal anomalies are $
c= \frac{29}{24}\,, a= \frac{13}  {16}$. To study the protected spectrum of the theory it is very useful to compute the supersymmetric index \cite{induec}. Using the standard definitions this is given as an expansion in the superconformal fugacities $q$ and $p$ \cite{Dolan:2008qi} as,
\be
1+\overline{{\bf 27}} h^{-1} (q p)^{\frac49} +h^3 (q p)^{\frac23}+ ... +(-{\bf 78} -1) q p+ ... \, .
\ee The bold-face numbers are representations of $E_6$ as we will elaborate momentarily, and $h$ is the fugacity for $U(1)_h$.
We remind the reader that the power of $q p$  is half the R charge for scalar operators  and we observe that all the operators are above the unitarity bound. 
 Let us count some of the  operators contributing to the index. The relevant operators of the model are $Q_B Q_A$ and $M_A$ which comprise the $({\bf 2},{\bf 6})$ and $({\bf 1},\overline {\bf 15})$    of $(SU(2)_B,SU(6)_A)$, which gives $\overline {\bf 27}$ of $E_6$. We also have $M_B$, a singlet of non abelian symmetries. At order $q \, p$, assuming the theory flows to an interacting conformal fixed point,  the index gets contributions only from marginal operators minus conserved currents for global symmetries \cite{Beem:2012yn}.
 The operators contributing at  order $q \, p$ are gaugino bilinear
$\lambda\lambda$ ($({\bf 1},{\bf 1})$), $Q_A \overline \psi_{Q_A}$ ($({\bf 1},{\bf 35}+{\bf 1})$), $Q_B\overline \psi_{Q_B}$ ($({\bf 3}+{\bf 1}, {\bf 1})$): these operators give the contribution, 
$$1-({\bf 1},{\bf 35})-1-({\bf 3},{\bf 1})-1,$$ which gives the conserved currents for the symmetry we see in the Lagrangian. Here $\overline \psi_{F}$ is the complex conjugate Weyl fermion in the chiral multiplet of the scalar $F$. We also have operators  $\overline \psi_{M_A} M_A$, $\overline \psi_{M_B} M_B$, $ M_B Q_B Q_B$, and $Q_A Q_A M_A$, which cancel out in the computation since the first two  are  fermionic  while the second two are bosonic, but are both in same representations of flavor symmetry pairwise. Finally we have $Q_A^3 Q_B$ ($({\bf 2},{\bf 70})$) and $Q_A \overline \psi_{M_A} Q_B$ ($({\bf 2}, {\bf 20}+{\bf 70})$). These two contribute $$-({\bf 2},{\bf 20})$$ to the index, which, combined with the above, forms the character of the adjoint representation of the $E_6\times U(1)_h$ symmetry. We emphasize that the fact we see $-{\bf 78}$ in order $q p$ of the index is a proof following from representation theory of the superconformal algebra that the symmetry of the theory enhances to $E_6$, where the only assumption is that the theory flows to interacting fixed point. We also observe that the conformal manifold here is a point.

\noindent{\bf Symmetry and duality: }  The enhancement of symmetry to $E_6$ follows from a well known IR duality \cite{Seiberg:1994pq}.  Note that we can reorganize the gauge charged  matter into two groups of four chiral fields. We take four out of the six $Q_A$ and call them $\widetilde Q$ and combine the other two with $Q_B$ and call those $Q$. This also decomposes the symmetry $SU(6)_A$ to $SU(4)_A \times SU(2)_A \times  U(1)_{h'}$ with a combination of  $U(1)_{h'}$ and $U(1)_h$ being the baryonic symmetry. See Fig. 2. IR duality  \cite{Seiberg:1994pq} without the gauge singlet fields will then map the baryonic symmetry to itself while conjugating the two $SU(4)$ symmetries and adding sixteen gauge singlet mesonic operators. With our choice of gauge singlet fields, the flipper fields of Fig. 2 are flipping eight of the baryons and the bifundamental gauge invariant operators form half of the mesons. Thus the duality removes the half of the mesons which connect $SU(2)_B$ with $SU(4)_A$ and attaches the other half between the $SU(4)_A$ and  $SU(2)_A$. This transformation acts only on the symmetry leaving the quiver structurally unchanged. The action on the symmetry is as the Weyl transformation which enhances $SU(6)_A\times  SU(2)_B$ symmetry to $E_6$.

\

\begin{figure}[h]
\includegraphics[scale=0.42]{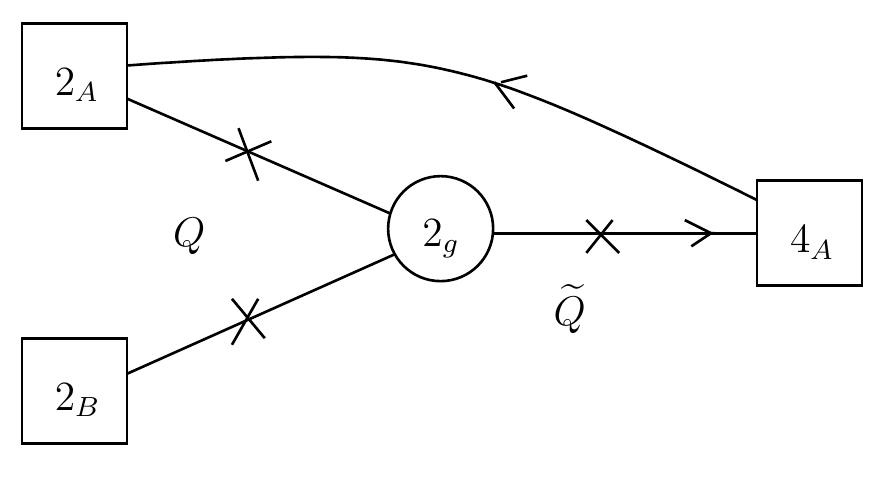}\caption{Equivalent way to represent the theory of Fig 1.}
\end{figure}

\noindent{\bf Deformation to Wess-Zumino (WZ) model: } Let us deform our model by adding $M_B$ to the superpotential which entails a vacuum expectation value for $Q_B Q_B$. This will Higgs the gauge group and we will be left with a WZ model of twenty seven chiral fields connected through a superpotential. See Fig. 3. In fact this model is the same as the one discussed in \cite{Razamat:2016gzx} (see for related observations  \cite{Gahramanov:2013xsa}). The superopotential can be thought of as a determinant of a three by three hermitian octonionic matrix, exactly the form of which the $E_6$ symmetry preserves. Although in four dimensions this is not too interesting of a model, all our arguments admit a generalization by reducing on a circle to three dimensions where this WZ model flows to interacting fixed point.

\begin{figure}[h]
\includegraphics[scale=0.46]{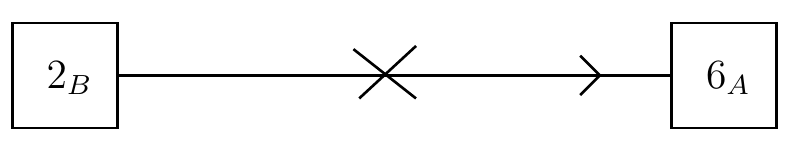} \caption{Model obtained deforming the quiver by admixing $M_B$ to the superpotential.}\end
{figure}

\noindent{\bf Relation to E string compactifications: }
The theory we have discussed can be related to compactifications of the E string, a superconformal theory in six dimensions with $E_8$ flavor symmetry. In Fig. 4 we depict the field theory one obtains by compactifying the rank one E string on a torus with half unit of flux breaking $E_8$ to $E_6\times U(1)$ \cite{Kim:2017toz}.
 Giving vacuum expectation value to the flip fields of this model gives us exactly the model of Fig 1. up to the singlet field $M_B$ which does not effect the enhancement of symmetry.  This relation provides a geometric explanation of the enhancement of symmetry. 

\begin{figure}[htb]
\includegraphics[scale=0.42]{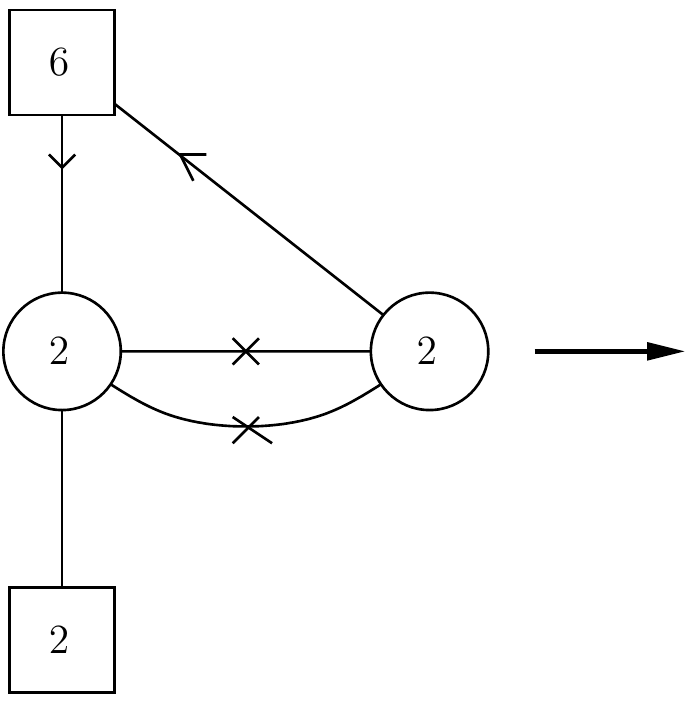}\includegraphics[scale=0.42]{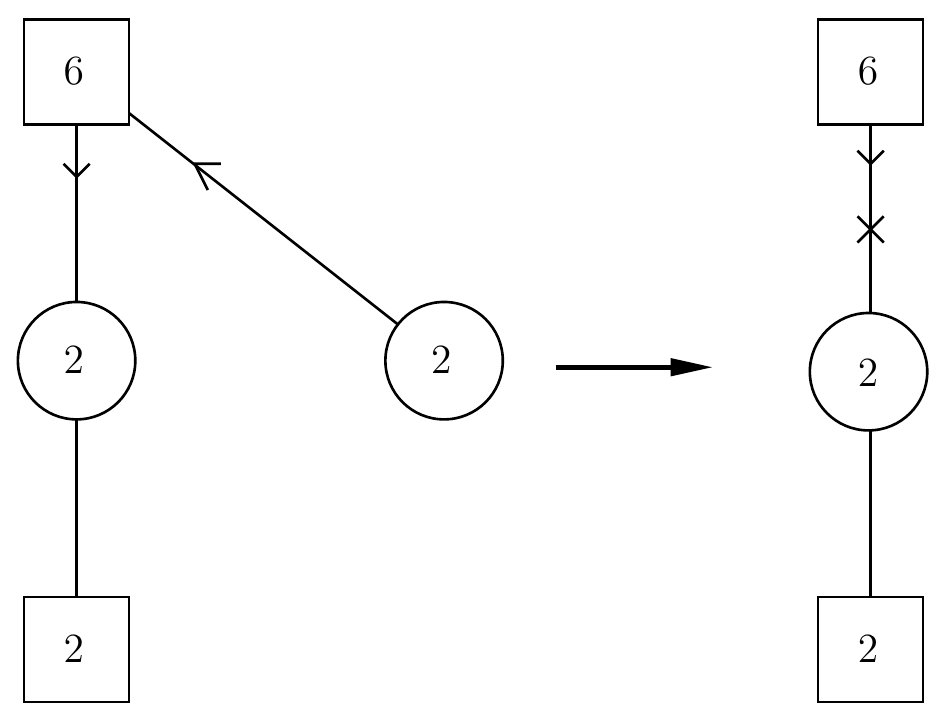}\caption{The model on the left corresponds to compactification of rank one E string on a torus with a particular flux  for the global symmetry. Giving vacuum expectation values to the flip fields one obtains the model in the middle which is equivalent under Seiberg duality to the third model.}
\end{figure}

We thus summarize that starting from $SU(2)_g$ gauge theory with four flavors and  adding gauge singlet fields breaking the $SU(8)$ symmetry to $U(1)_h\times SU(6)_A\times SU(2)_B$, the RG flow leads to a model with $E_6\times  U(1)_h$ symmetry. This statement only assumes that the flow leads to interacting fixed points and we see no evidence to the contrary. The enhancement of symmetry discussed here is very closely related to enhancement of symmetry of the sequence of $USp(4n)_g$ gauge theories discussed in \cite{Razamat:2017hda}  following \cite{Dimofte:2012pd}.

\section{Duality from $SO(12)$  symmetry}

The second case we study is  $SU(4)_g$ gauge theory with four flavors, $Q$ and $\widetilde Q$, and two chiral fields, $X$,  in the antisymmetric representation. The fields and the charges are  in the upper half of the table below.

\begin{tabular}{|c||c|c|c|c|c|c|c|}
Field&$SU(4)_g$&$SU(4)$&$SU(4)$&$SU(2)$&$U(1)_a$&$U(1)_b$&$U(1)_{r}$\\ \hline
 $Q$&${\bf 4}$&${\bf 4}$ &${\bf 1}$&${\bf 1}$& $-1$&$\,\;1$&$\frac12$ \\
$\widetilde Q$&$\overline {\bf 4}$&${\bf 1}$&${\bf 4}$&${\bf 1}$&$-1$&$-1$&$\frac12$\\ 
$X$&${\bf 6}$&${\bf 1}$&${\bf 1}$&${\bf 2}$&$\; \, 2$&$\;\,0$&$0$\\ \hline\hline 
$M$&${\bf 1}$ &$\overline {\bf 4}$ & $\overline {\bf 4}$ & $ {\bf 1} $ &  $-2$ & $\;\,0$ & $1$\\
$B$ & ${\bf 1}$ & $ {\bf 6}$ & ${\bf 1}$ & ${\bf 2}$ & $\,\;0$ & $-2$ & $1$\\
$\phi$ & ${\bf 1}$ & ${\bf 1}$ &${\bf 1}$ & ${\bf 3} $ &  $-4$ & $\;\,0$ & $2$\\
\end{tabular}

\

This model and the one discussed in the previous section are two first entries, $N=2,\, 1$, in a sequence of $SU(2N)_g$ gauge theories with four flavors in fundamental representation  and a field in antisymmetric representation. All models in this sequence are known to have self dual descriptions \cite{Csaki:1997cu, Spiridonov:2009za}. We will not detail the dualities here, just mention that  the two  $SU(4)$ symmetry groups are manifest in all dual frames. 
We can construct a model with $SU(4)_g$ gauge symmetry which is dual to itself by turning on a very particular collection of gauge singlet fields listed in the bottom half  of the table above. We couple the gauge singlet fields with the following superpotential,
\be 
\widetilde Q Q X^2 M+ Q Q X B +\phi X^2\,.
\ee As all the dualities preserve the global symmetry and the superpotential also does not  break it we naively do not expect any non abelian enhancement of symmetry following from these dualities. However computing the index we find that all the states fall into representations of $SO(12) \times SU(2)\times U(1)^2$,
\be
&&
1+({\bf 1}, {\bf 32}') a^{-2} (q p)^{\frac13} +({\bf 2},{{\bf 12}})b^{-2} (q p)^{\frac12}+\dots \\ && +(-({\bf 1},{\bf 66})-({\bf 3} ,{\bf 1})-1-1) q p + . . .    \;.
\nonumber \ee Here $({\bf {\frak R}_1},{\bf {\frak R}_2})$  denotes characters of representation ${\bf {\frak R}_1}\times   
{\bf \frak R}_2$ of $SU(2)  \times SO(12)$, and $a, b$ are fugacities for $U(1)_a, U(1)_b$ respectively.
 The superconformal R symmetry here is given as $$\hat r  = r+0.057 {\frak q}_b+0.142  {\frak q}_a$$ and we used to evaluate the index the R charge $r+\frac16{\frak q}_a$ which is close to the superconformal one. The two $SU(4)$ symmetry groups  are imbedded in $SO(12)$ as  $SO(6)\times SO(6)$. Moreover, at the order of the index in which the conserved currents and marginal deformations contribute we observe a term in adjoint of $SO(12)$ indicating that the symmetry enhances to this in the IR.  The decomposition into $SO(6)\times SO(6)$ representations of the adjoint is,
\be {\bf 66}= ({\bf 15}, {\bf 1})+({\bf 1}, {\bf 15})+({\bf 6}, {\bf 6})\,.
\ee  The first two terms come from $Q\overline \psi_Q$ and $\widetilde Q\overline{ \psi}_{\widetilde Q}$. We have an operator in $({\bf 6} ,  {\bf 6})$ residing in the bosonic operator $\widetilde Q^2 Q^2 X^2$, and two fermionic operators, $\widetilde Q^2 \overline \psi_B X$ with $\widetilde Q Q \overline \psi_M$. The combined effect of these is to contribute $({\bf 6},\, {\bf 6})$ to the enhancement of the conserved current.

\noindent{{\bf Duality from Symmetry: }  The enhancement to $SO(12)$ does not follow from the dualities of \cite{Csaki:1997cu} as these preserve the two $SU(4)$ symmetries. However the Weyl group for $SO(12)$ permutes Cartan generators residing in the two $SU(4)$ symmetries. Enhancement of symmetry can be taken as an indication that there is a dual description of the original theory in which the two $SU(4)$ symmetries are broken and the subgroups are identified in a non obvious way.  The basic duality of \cite{Csaki:1997cu} is a generalization of Seiberg duality for $SU(2)_g$ theory with eight flavors. There we have $35$ different dualities corresponding to splitting the eight chiral fields into two groups of four. 
 However here we naively do not have a generalization of such freedom as the relevant representations are complex. The enhancement of symmetry is again an indication that such a generalization can be obtained as we shall now describe.

We first decompose both $SU(4)$ symmetries into $SU(2)\times SU(2)\times U(1)$ and write the matter content of the $SU(4)_g$ theory with no gauge singlet fields in terms of representations of these groups. In the table we detail the charges under the $SU(4)$ symmetries as the rest are as in the previous table.

\begin{tabular}{|c||c|c|c|c|c|c|}
Field&$SU(2)_L$&$\widetilde {SU}(2)_L$&$U(1)_L$&$SU(2)_R$&$\widetilde {SU}(2)_R$& $U(1)_R$\\ \hline
 $Q^-$&${\bf 2}$ &${\bf 1}$&$-1$& ${\bf 1}$&${\bf 1}$&$\;\,0$ \\
 $Q^+$&${\bf 1}$ &${\bf 2}$&$\;\,1$& ${\bf 1}$&${\bf 1}$&$\;\,0$ \\
$\widetilde Q^-$&${\bf 1}$&${\bf 1}$&$\;\,0$&${\bf 2}$&${\bf 1}$&$-1$\\ 
$\widetilde Q^+$&${\bf 1}$&${\bf 1}$&$\;\,0$&${\bf 1}$&${\bf 2}$&$\;\,1$\\ 
$X$&${\bf 1}$&${\bf 1}$&$\;\,0$&${\bf 1}$&${\bf 1}$&$\;\,0$\\ \hline

\end{tabular}

\

The dual theory has the following fields, 

\

\begin{tabular}{|c||c|c|c|c|c|c|}
Field&$SU(2)_L$&$\widetilde {SU}(2)_L$&$U(1)_L$&$SU(2)_R$&$\widetilde {SU}(2)_R$& $U(1)_R$\\ \hline
 ${\frak Q}^-$&${\bf 2}$ &${\bf 1}$&$\;\,0$& ${\bf 1}$&${\bf 1}$&$-1$ \\
 ${\frak Q}^+$&${\bf 1}$ &${\bf 2}$&$\;\,0$& ${\bf 1}$&${\bf 1}$&$\;\,1$ \\
$\widetilde {\frak Q}^-$&${\bf 1}$&${\bf 1}$&$-1$&${\bf 2}$&${\bf 1}$&$\;\,0$\\ 
$\widetilde {\frak  
Q}^+$&${\bf 1}$&${\bf 1}$&$\;\,1$&${\bf 1}$&${\bf 2}$&$0$\\ 
$\frak X$&${\bf 1}$&${\bf 1}$&$\;\,0$&${\bf 1}$&${\bf 1}$&$0$\\ \hline\hline 
${\frak M}^{-+}_l$&${\bf 1}$ & ${\bf 2}$ & $ \; \,1$ &  ${\bf 2}$ & ${\bf 1}$ & $-1$\\
${\frak M}^{+-}_l$&${\bf 2}$ & ${\bf 1}$ & $ -1 $ &  ${\bf 1}$ & ${\bf 2}$ & $\;\,1$\\
${\frak B}^{\mp}$ &  ${\bf 1}$ & ${\bf 1}$ & $\;\,0$ & ${\bf 1}$ & ${\bf 1}$ & $\mp2$\\
${\widetilde{\frak B}}^{\mp}$&${\bf 1}$ & ${\bf 1}$ & $\mp2$ & ${\bf 1}$ & ${\bf 1}$ & $\;\,0$\\  \hline

\end{tabular}

\

The duality exchanges the $U(1)_L$ and $U(1)_R$ symmetries. The superpotential is,
\be &&
\sum_{l=0}^1({\frak M}^{-+}_l {\frak Q}^+ {\widetilde {\frak Q}}^- +{\frak M}^{+-}_l {\frak Q}^-{\widetilde {\frak Q}}^+){\frak X}^{2(1-l)}+\\ && \;\,({\frak B}^+ {\frak Q}^- {\frak Q}^-+{\widetilde {\frak B}}^+ {\widetilde {\frak Q}}^- {\widetilde {\frak Q}}^-){\frak X}+({\frak B}^- {\frak Q}^+ {\frak Q}^++{\widetilde{\frak B}}^- {\widetilde {\frak Q}}^+ {\widetilde {\frak Q}}^+){\frak X}\,.\nonumber
\ee
We can make many checks on this duality, in particular  that the indices of the two dual models are in agreement and that the 't Hooft anomalies match.  The contribution of the singlet fields to all the abelian anomalies is vanishing. The non obvious matching of anomalies involves the $SU(2)$ symmetries.
For example,
\be
Tr SU(2)_L^2U(1)_L&\,& = \left[{Q^-}\right] \\
&\;&= \left[{\frak M}_1^{+-}\right]+\left[{\frak M}_2^{+-}\right]  = -2\, .\nonumber
\ee

\

Let us mention that relevant deformation of the $SO(12)$  theory with $Q^4$ operator leads to symmetry enhancing to $E_7\times U(1)$.

\
 
We summarize by stating that using the basic observations of this note one can generate many examples of conformal theories with enhanced symmetries starting from known self dualities and derive new self dualities by observing enhancements of symmetry. 
We plan to report in more detail on some of the plethora of examples in a forthcoming work.

\

\noindent{\bf Acknowledgments}:~
GZ is supported in part by  World Premier International Research Center Initiative (WPI), MEXT, Japan.  The research of SSR and of OS was  supported by Israel Science Foundation under grant no. 1696/15 and by I-CORE  Program of the Planning and Budgeting Committee.

\end{document}